\documentclass[aip,jcp,reprint,amsmath,amssymb,amsfonts,unsortedaddress]{revtex4-1}
\usepackage{graphicx,dcolumn,bm,color,microtype,multirow,mathtools,setspace,hyperref,subfigure,tabularx,pifont}

\newcommand{\alert}[1]{\textcolor{black}{#1}}

\begin{document}

\title{Nodal surfaces and interdimensional degeneracies}

\author{Pierre-Fran{\c c}ois Loos}
\email{Corresponding author: pf.loos@anu.edu.au}
\affiliation{Research School of Chemistry, Australian National University, Canberra ACT 2601, Australia}
\author{Dario Bressanini}
\email{dario.bressanini@uninsubria.it}
\affiliation{Dipartimento di Scienza e Alta Tecnologia, Universit\`a dell'Insubria, Via Lucini 3, I-22100 Como, Italy}

\begin{abstract}
The aim of this paper is to shed light on the topology and properties of the nodes (i.e. the zeros of the wave function) in electronic systems.
Using the ``electrons on a sphere'' model, we study the nodes of two-,  three- and four-electron systems in various ferromagnetic configurations ($sp$, $p^2$, $sd$, $pd$, $p^3$, $sp^2$ and  $sp^3$).
In some particular cases ($sp$, $p^2$, $sd$, $pd$ and $p^3$), we rigorously prove that the non-interacting wave function has the same nodes as the exact (yet unknown) wave function.
The number of atomic and molecular systems for which the exact nodes are known analytically is very limited and we show here that this peculiar feature can be attributed to interdimensional degeneracies.
Although we have not been able to prove it rigorously, we conjecture that the nodes of the non-interacting wave function for the $sp^3$ configuration are exact.
\end{abstract}

\keywords{Fermion nodes; full configuration interaction; quantum Monte Carlo; fixed-node error}
\pacs{}
	
\maketitle

\section{Introduction}
\alert{Considering an antisymmetric (real) electronic wave function} $\Psi(\bm{\mathcal{S}},\bm{R})$, where $\bm{\mathcal{S}} = (s_1,s_2,\ldots,s_n)$  are the spin coordinates and $\bm{R} = (\bm{r}_1,\bm{r}_2,\ldots,\bm{r}_n)$ are the $D$-dimensional spatial coordinates of the $n$ electrons, the nodal hypersurface (or simply ``nodes'') is a $(n\, D-1)$-dimensional manifold defined by the set of configuration points $\bm{N}$ for which $\Psi(\bm{N})=0$.
The nodes divide the configuration space into nodal cells or domains which are either positive or negative depending on the sign of the electronic wave function in each of these domains.
In recent years, strong evidence has been gathered showing that, for the lowest state of any given symmetry, there is a single nodal hypersurface that divides configuration space into only two nodal domains (one positive and one negative). \cite{Ceperley91, Glauser92, Bressanini01, Bressanini05a, Bajdich05, Bressanini05b, Scott07, Mitas06, Mitas08, Bressanini08, Bressanini12}
In other words, to have any chance to have \textit{exact} nodes, a wave function must have only two nodal cells. 
For simplicity, in the remainder of this paper, we will say that a wave function has exact nodes if it has the same nodes as the exact wave function.
\alert{Except in some particular cases, electronic or more generally fermionic nodes are poorly understood due to their high dimensionality and complex topology. \cite{Ceperley91, Bajdich05}}
The number of systems for which the exact nodes are known analytically is very limited. 
For atoms, it includes two triplet --- $^3S^\text{e}(1s2s)$ and $^3P^\text{e}(2p^2)$ --- and two singlet --- $^1S^\text{e}(2s^2)$ and $^1P^\text{e}(2p^2)$ --- states of the helium atom and the three-electron atomic state $^4S(2p^3)$. \cite{Klein76, Bressanini05a, Bajdich05}

The quality of fermion nodes is of prime importance in quantum Monte Carlo (QMC) calculations due to the fermion sign problem, which continues to preclude the application of in principle exact QMC methods to large systems.
The dependence of the diffusion Monte Carlo (DMC) energy on the quality of the trial wave function is often significant in practice, and is due to the fixed-node approximation which segregates the walkers in regions defined by the trial or guiding wave function.
\alert{The fixed-node error is only proportional to the square of the nodal displacement error, but it is uncontrolled and its accuracy difficult to assess. \cite{Kwon98, Luchow07a, Luchow07b}}
Recently, Mitas and coworkers have discovered a interesting relationship between electronic density, degree of node nonlinearity and fixed-node error. \cite{Kulahlioglu14, Rasch12, Rasch14}

Here, we study the topology and properties of the nodes in a class of systems composed of electrons located on the surface of a sphere.
Due to their high symmetry and their mathematical simplicity, these systems are the ideal ``laboratory'' to study nodal hypersurface topologies in electronic states.
Moreover, Mitas showed that the non-interacting wave function of spin-polarized electrons on a sphere has only two nodal cells which is probably a necessary condition for exactness \cite{Mitas06, Mitas08} (see above).
\alert{Although the present paradigm can be seen as over simplified, it has been successfully used to shed light on the adiabatic connection within density-functional theory, \cite{Seidl07} to prove the universality of the correlation energy of an electron pair \cite{EcLimit09, Ballium10, EcProof10, Frontiers10}, as a model for ring-shaped semiconductors (known as quantum rings), \cite{QR12} to understand the properties of excitons, \cite{Exciton12} and to create finite \cite{Glomium11, Ringium13} and infinite \cite{1DEG13, 2DEG11, 3DEG11} uniform electron gases \cite{UEGs12} and new correlation density-functionals. \cite{GLDA1, GLDAw}}

In this paper, we report the analytic expression of the exact nodes for two-electron triplet states ($sp$, $p^2$, $sd$ and $pd$). 
We also show that, as in the atomic case, the nodes of the non-interacting wave function for the three-electron state $^4S(p^3)$ are identical to the nodes of the exact wave function. 
In addition to these systems where the non-interacting wave function has exact nodes, we study the quality of the non-interacting nodes for the $sp^2$ and $sp^3$ configurations. 
For the $sp^2$ configuration, we show that, although not exact, the non-interacting nodes are very accurate and, based on numerical evidences, we conjecture that the non-interacting and exact nodes of the $sp^3$ configuration are identical.
We use atomic units throughout. 

\begin{table*}
	\caption{
	\label{tab:HFwf}
	Non-interacting wave function $\Psi_0$ for various ferromagnetic states of $n$ electrons on a sphere and their corresponding irreducible representation (IR) in $\text{D}_\text{2h}$.
	$\bm{z}=(0,0,1)$ is the unit vector of the $z$ axis, $\bm{r}_{ij} = \bm{r}_i - \bm{r}_j$, $\bm{r}_{ij}^+ = \bm{r}_i + \bm{r}_j$ and $\bm{r}_{ij}^\times = \bm{r}_i \times \bm{r}_j$.}
		\begin{ruledtabular}
			\begin{tabular}{ccccccccc}
				$n$		&	State				&	Configuration	&	$\Psi_0(\{\bm{\Omega}_i\})$								&	$\text{D}_\text{2h}$ IR																			& 	Exact?	\\
				\hline
				2		&	$^3P^{\rm o}$		&	$sp$			&	$\bm{z} \cdot \bm{r}_{12}$										&	$\text{A}_\text{g} \otimes \text{B}_\text{1u} =  \text{B}_\text{1u}$										&	Yes	\\
				2		&	$^3P^{\rm e}$		&	$p^2$		&	$\bm{z} \cdot \bm{r}_{12}^\times$								&	$\text{B}_\text{3u} \otimes \text{B}_\text{2u} =  \text{B}_\text{1g}$										&	Yes	\\
				2		&	$^3D^{\rm e}$		&	$sd$			&	$(\bm{z} \cdot \bm{r}_{12}^+)(\bm{z} \cdot \bm{r}_{12})$				&	$\text{A}_\text{g} \otimes \text{A}_\text{g} =  \text{A}_\text{g}$											&	Yes	\\
				2		&	$^3D^{\rm o}$		&	$pd$			&	$(\bm{z} \cdot \bm{r}_{12}^+)(\bm{z} \cdot \bm{r}_{12}^\times)$			&	$\text{B}_\text{2g} \otimes \text{B}_\text{2u} = \text{B}_\text{3g} \otimes \text{B}_\text{3u} = \text{A}_\text{u}$		&	Yes	\\
				3		&	$^4S^{\rm o}$		&	$p^3$		&	$\bm{r}_1 \cdot \bm{r}_{23}^\times$								&	$\text{B}_\text{1u} \otimes \text{B}_\text{2u}  \otimes \text{B}_\text{3u} = \text{A}_\text{u}$						&	Yes	\\
				3		&	$^4D^{\rm e}$		&	$sp^2$		&	$\bm{z} \cdot (\bm{r}_{12} \times \bm{r}_{13})$						&	$\text{A}_\text{g} \otimes \text{B}_\text{3u} \otimes \text{B}_\text{2u} =  \text{B}_\text{1g}$						&	No	\\
				4		&	$^5S^{\rm o}$		&	$sp^3$		&	$(\bm{r}_{12} + \bm{r}_{34})(\bm{r}_{12}^\times + \bm{r}_{34}^\times)$	&	$\text{A}_\text{g} \otimes \text{B}_\text{1u} \otimes \text{B}_\text{2u}  \otimes \text{B}_\text{3u} = \text{A}_\text{u}$	&	Unknown	\\
			\end{tabular}
		\end{ruledtabular}
\end{table*}

\section{Electrons on a sphere}
Our model consists of $n$ spin-up electrons restricted to remain a surface of a sphere. \cite{TEOAS09, Glomium11}
The non-interacting orbitals for an electron on a sphere of radius $R$ are the normalized spherical harmonics $Y_{\ell m}(\bm{\Omega})$, where $\bm{\Omega} =(\theta,\phi)$ are the polar and azimuthal angles respectively.
We will label spherical harmonics with $\ell=0,1,2,3,4,\ldots$ as $s$, $p$, $d$, $f$, $g$, \ldots functions.
The coordinates of the electrons on the unit sphere are given by their cartesian coordinates $x = \cos \phi \sin \theta$, $y = \sin \phi  \sin \theta$ and $z = \cos \theta$.
The average electronic density is measured by the so-called Wigner-Seitz radius $r_s = (\sqrt{n}/2)\,R$. 

The Hamiltonian of the system is simply
\begin{equation}
	\Hat{H} = -\frac{1}{2} \sum_i^n \nabla_i^2 + \sum_{i<j}^n \frac{1}{r_{ij}},
\end{equation}
where $\nabla_i^2$ is the angular part of the Laplace operator for electron $i$ and $r_{ij} = \left| \bm{r}_i - \bm{r}_j\right|$ is the interelectronic distance between electrons $i$ and $j$, i.e. the electrons interact \emph{through} the sphere.
\alert{\footnote{\alert{Roughly speaking, our model can be viewed as an atom in which one only considers the subspace $r_1 = r_2 = \ldots = r_n = R$.}}}
We write the electronic wave function as
\begin{equation}
\label{Phi}
	\Phi (\{s_i\},\{\bm{\Omega}_i\})
	= \Xi(\{s_i\})\,\Psi_0(\{\bm{\Omega}_i\})\,\Lambda(\{\bm{\Omega}_i\}).
\end{equation}
$\Xi$ is the spin wave function and only depends on the spin coordinates $\{s_i\}$. 
Because we only consider ferromagnetic systems, the spin wave function is $\Xi(\{s_i\}) = \prod_{i=1}^n \alpha(s_i)$, and is symmetric with respect to the interchange of two electrons. 
The non-interacting wave function  $\Psi_0$ is a Slater determinant of spin orbitals and defines the nodal hypersurface.
$\Lambda$ is a \emph{nodeless} correlation factor and, because $\Psi_0$ is antisymmetric, this means that $\Lambda$ has to be symmetric with respect to the exchange of two electrons.

We will label each state using the following notations: $^{2\mathcal{S}+1}L^{\rm e,o}$, where $L=S,P,D,F,\ldots$ and $\mathcal{S} = \sum_{i=1}^n s_i$ is the total spin angular momentum. 
The suffixes e (even) and o (odd) are related to the parity of the states given by $(-1)^{\ell_1+\cdots+\ell_n}$.

\begin{figure*}
	\subfigure[$^3P^\text{o}(sp)$]{
		\includegraphics[width=0.22\textwidth]{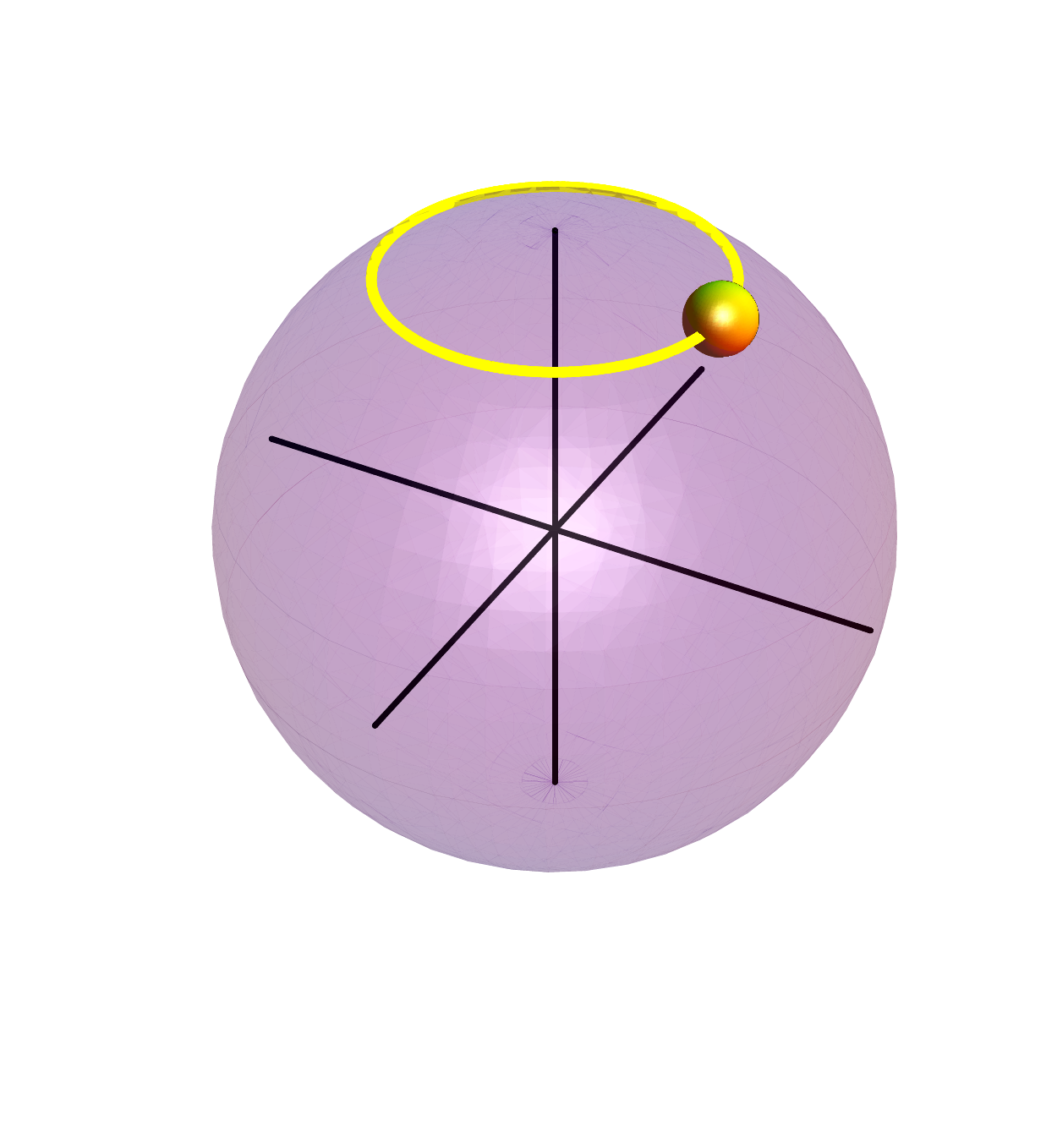}
	}
	\subfigure[$^3P^\text{e}(p^2)$]{
		\includegraphics[width=0.22\textwidth]{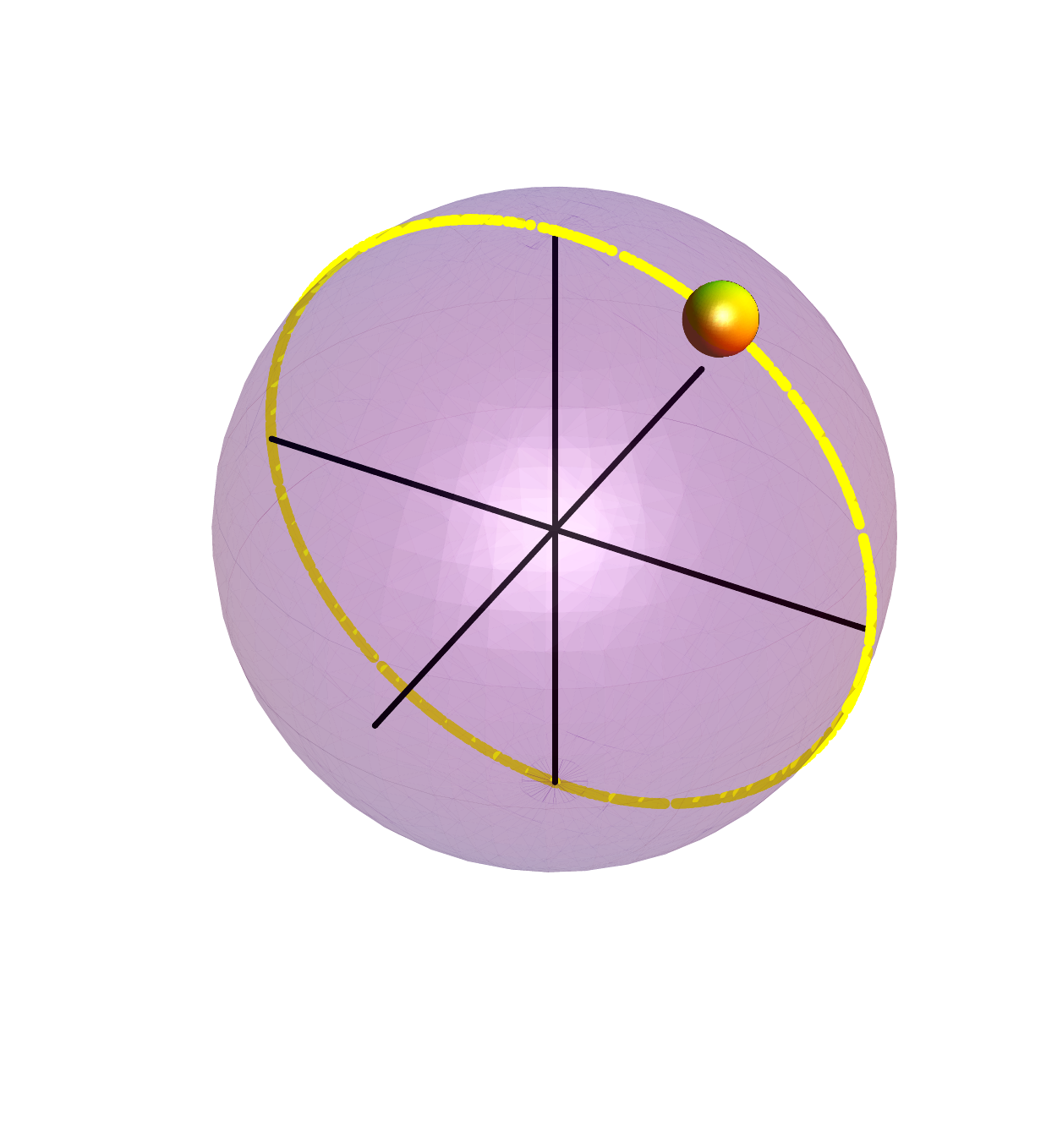}
	}
	\subfigure[$^3D^\text{e}(sd)$]{
		\includegraphics[width=0.22\textwidth]{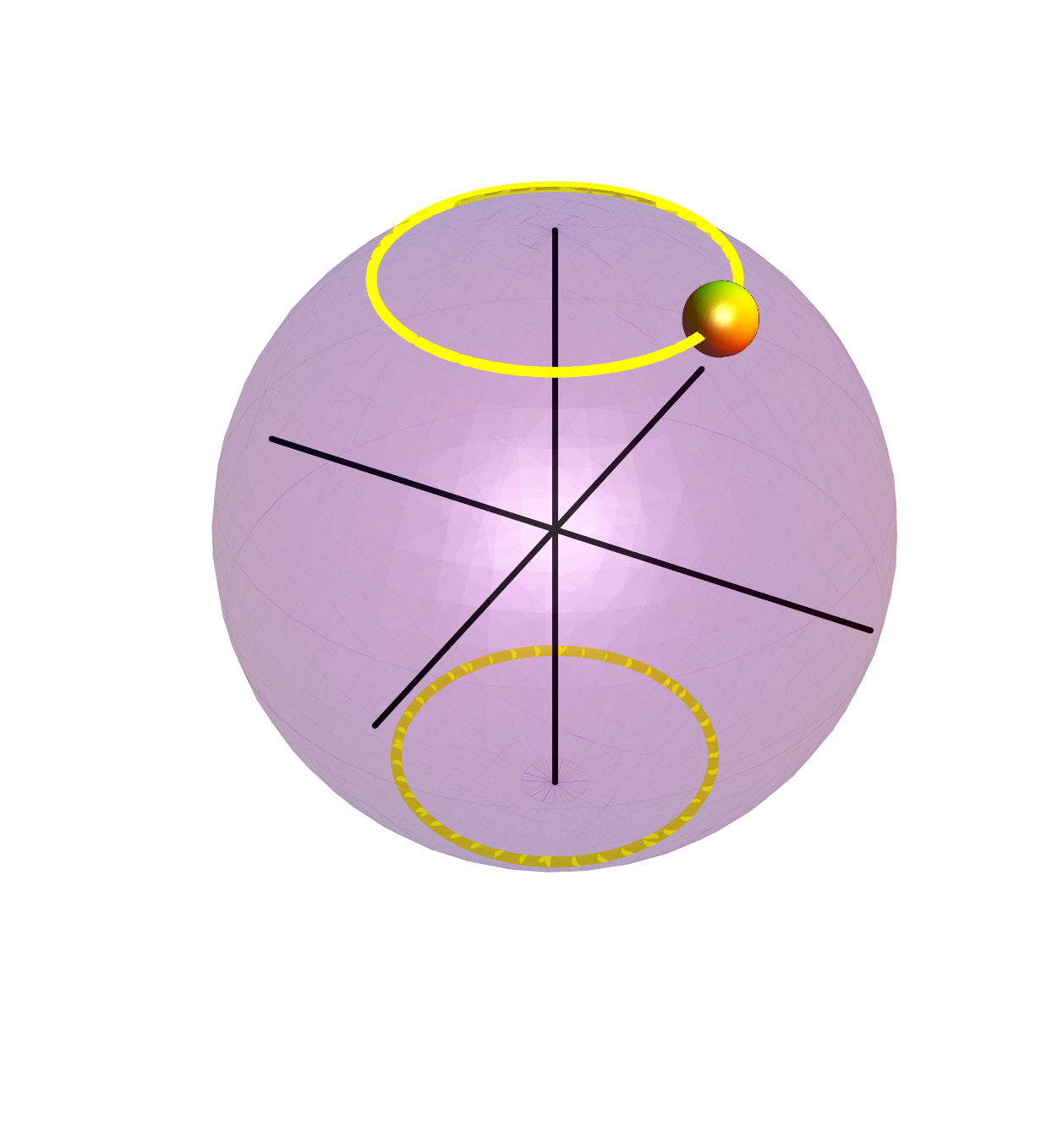}
	}
	\subfigure[$^3D^\text{o}(pd)$]{
		\includegraphics[width=0.22\textwidth]{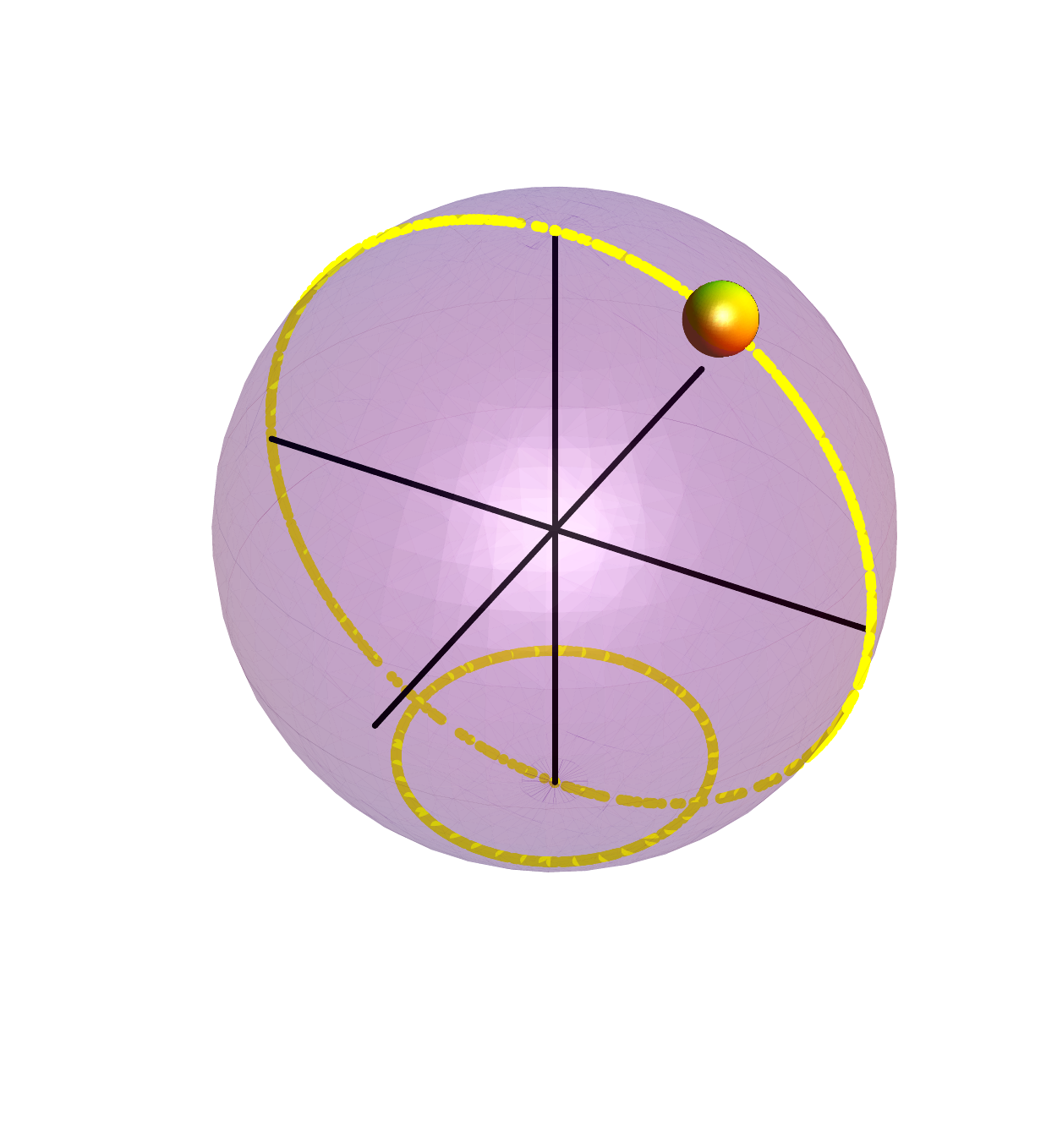}
	}
	\caption{
	\label{fig:2e}
	Non-interacting nodes for various two-electron ferromagnetic states. 
	The first electron is at $\bm{\Omega}_1=(\pi/6,0)$ (large yellow dot).
	The possible position of the second electron corresponding to $\Psi_0 = 0$ are represented by \alert{a yellow line}.
	}
\end{figure*}

\subsection{Two-electron systems}

\subsubsection{Non-interacting wave functions}
First, we study ferromagnetic two-electron (i.e. triplet) states.
For each two-electron state gathered in Table \ref{tab:HFwf}, the non-interacting wave function takes a simple form:
\begin{align}
	\label{sp-nodes}
	\Psi_0(sp) & = 
		\begin{vmatrix}
			1	&	z_1		\\
			1	&	z_2		\\
		\end{vmatrix}
	= \bm{z} \cdot \bm{r}_{12},
	\\
	\label{p2-nodes}
	\Psi_0(p^2) & = 
		\begin{vmatrix}
			x_1	&	y_1		\\
			x_2	&	y_2		\\
		\end{vmatrix}
	= \bm{z} \cdot \bm{r}_{12}^\times,
	\\
		\Psi_0(sd) & = 
		\begin{vmatrix}
			1	&	2z_1^2-x_1^2-y_1^2		\\
			1	&	2z_2^2-x_2^2-y_2^2		\\
		\end{vmatrix}
	=(\bm{z} \cdot \bm{r}_{12}^+)(\bm{z} \cdot \bm{r}_{12}),
	\\
	\Psi_0(pd) & = 
		\begin{vmatrix}
			y_1	&	x_1 z_1		\\
			y_2	&	x_2 z_2		\\
		\end{vmatrix}
		-
				\begin{vmatrix}
			x_1	&	y_1 z_1		\\
			x_2	&	y_2 z_2		\\
		\end{vmatrix}
	=(\bm{z} \cdot \bm{r}_{12}^+)(\bm{z} \cdot \bm{r}_{12}^\times),
	\label{pd-nodes}
\end{align}
where $\bm{z}=(0,0,1)$ is the unit vector of the $z$ axis, $\bm{r}_{ij} = \bm{r}_i - \bm{r}_j$, $\bm{r}_{ij}^+ = \bm{r}_i + \bm{r}_j$ and $\bm{r}_{ij}^\times = \bm{r}_i \times \bm{r}_j$.
Due to their ferromagnetic nature, each state has ``Pauli'' nodes which corresponds to configurations where two electrons touch.
The Pauli hyperplanes are only a subset of the full nodes and it is interesting to note that these nodes are related to the 1D nodes. \cite{QR12, Ringium13, GLDA1}
The non-interacting nodes are represented in Fig.~\ref{fig:2e} for a given position of the first electron $\bm{\Omega}_1=(\pi/6,0)$, which is represented by a small yellow sphere.
The possible positions of the second electron for which $\Psi_0$ vanishes are represented by smaller yellow dots.

\subsubsection{Proof of the exactness of the nodes}
\begin{figure*}
	\begin{tabularx}{\linewidth}{|XcX|XcXcX|}	
	\hline
	\includegraphics[width=0.14\textwidth]{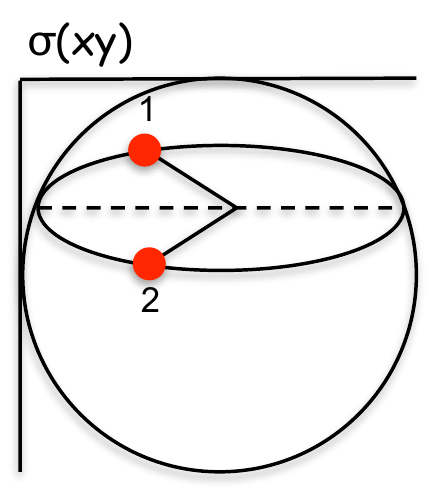}
	&
	$\xrightarrow[\Psi = \Psi^\prime]{\sigma(xy)}$
	&
	\includegraphics[width=0.135\textwidth]{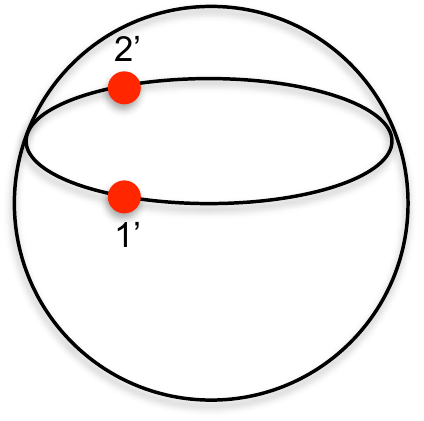}
	&
	\includegraphics[width=0.15\textwidth]{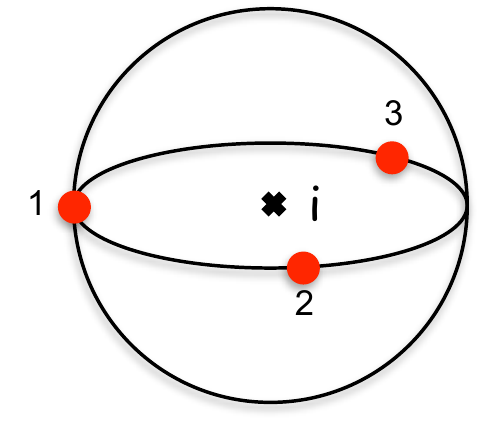}
	&
	$\xrightarrow[\Psi = -\Psi^\prime]{i}$
	&
	\includegraphics[width=0.15\textwidth]{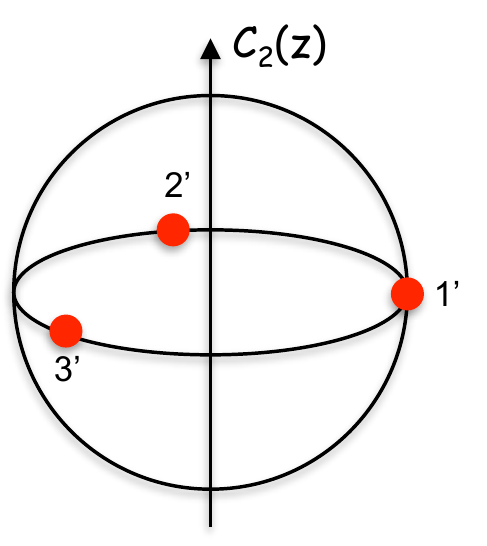}
	&
	$\xrightarrow[\Psi^\prime = \Psi^{\prime\prime}]{C_2(z)}$
	&
	\includegraphics[width=0.15\textwidth]{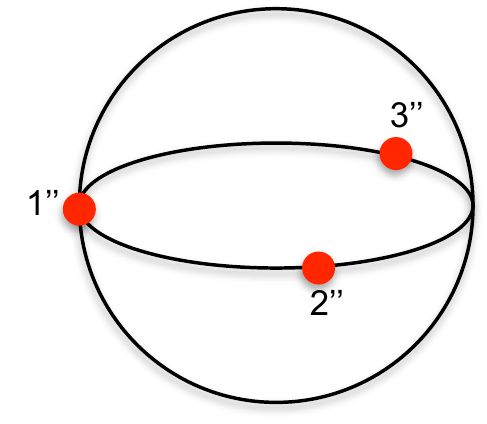}
	\\
	\hline
	\end{tabularx}
	\caption{
	\label{fig:proof}
	Proof of the exactness of the non-interacting nodes for the $^3P^\text{o}(sp)$ (left) $^4S^\text{o}(p^3)$ (right) states.
	}
\end{figure*}

Equation \eqref{sp-nodes} shows that the non-interacting nodes of the $sp$ configuration corresponds to small circle perpendicular to the $z$ axis.
Now, let us prove that these non-interacting and exact nodes are identical.
We begin by placing the two electrons on a small circle perpendicular to the $z$ axis, as sketched in Fig.~\ref{fig:proof}. 
For this particular configuration, the two electrons have the same value of the polar angle $\theta = \theta_1 = \theta_2$ and, without loss of generality, the azimuthal angles can be chosen such that $\phi_1 = - \phi_2 = \phi$. 
Suppose that for this configuration the exact wave function has a value $\Psi \equiv \Psi(\{(\theta,+\phi),(\theta,-\phi)\}) = K$. 
Now, we reflect the wave function with respect to the symmetry plane $\sigma(xz)$ that passes through the $x$ and $z$ axes and bisects the azimuthal angle $\phi$. 
Due to the $P$ nature of the state ($\text{B}_\text{1g}$ representation of the $\text{D}_\text{2h}$ point group as shown in Table \ref{tab:HFwf}), the wave function is invariant to such reflexion, i.e. $\Psi^\prime \equiv \Psi(\{(\theta,-\phi),(\theta,+\phi)\}) = K$. 
However, the two electrons have been exchanged and because this is a triplet state, the wave function must have changed sign (Pauli principle). 
Because $\Psi = - \Psi^\prime$, this implies that $K=-K$ which means that $K=0$ and $\forall (\theta,\phi), \Psi(\theta,\theta,\phi,-\phi) = 0$. 
We have discovered the nodes of the $sp$ configuration by using symmetry operators belonging to the symmetry group of this particular state.
This methodology can be applied to the other two-electron states to show that, in each case, the non-interacting nodes are the same as the nodes of the exact wave function.
We have confirmed these results by performing full configuration interaction (FCI) calculations \cite{Knowles84}, as well as near-exact Hylleraas-type calculations. 
For each of these calculations, we have shown that the non-interacting nodes never move when electronic correlation is taken into account. 
This provides a complementary ``computational'' proof of the exactness of the non-interacting nodes.
\alert{This observation also means that, for any size of the basis set, the FCI nodes are exact.}

However, we must show that these are the only nodes since there might be possibly be other nodal surfaces. 
For the ground state, the number of nodal cells is minimal and Mitas \cite{Mitas08} has shown that these systems have the minimal number of two nodal pockets.
As explained by Bajdich {\em et al.} \cite{Bajdich05}, any distortion of the node from the great circle leads to additional cells which can only increase energy by imposing higher curvature (kinetic energy) on the wave function. 
This argument has been used by Feynman to demonstrate that the energy of fermionic ground state is always higher than the energy of the bosonic state, and by Ceperley \cite{Ceperley91} to demonstrate the tiling property of the nodal surface. \footnote{The tiling theorem states that, for a non-degenerate ground state, by applying all possible particle permutations to an arbitrary nodal cell one covers the complete configuration space.}

It is interesting to note that the exact nodes of the $pd$ configuration can be represented using two Slater determinants --- see Eq.~\eqref{pd-nodes} --- and the nodal surface is composed of two intersecting nodal surfaces as shown in Fig.~\ref{fig:2e}.
This is in agreement with the result of Pechukas who showed that, when two nodal surfaces cross, they are perpendicular at the crossing point. \cite{Pechukas72}

We have recently shown that, for certain states such as the $^3P^{\rm o}(sp)$, $^3P^{\rm e}(p^2)$, $^3D^{\rm e}(sd)$ and $^3D^{\rm o}(pd)$ states, exact solutions of the Schr\"odinger equation can be found in closed form for specific values of the radius $R$. \cite{QuasiExact09, ExSpherium10}
Even though the exact closed-form expression of the Schr\"odinger equation is only known for particular values of the radius, their exact nodes are analytically known for all radii (see Table \ref{tab:HFwf}).

One could ask if there are any other two-electron states for which we know the exact nodes? The answer is no.
Each of the states having exact nodes is the lowest-energy state of a given irreducible representation of the $\text{D}_\text{2h}$ point group (the largest abelian point group in 3D).
For example, the states $^3D^\text{e}(sd)$ and $^3D^\text{o}(pd)$ correspond to the lowest-energy state of the $\text{A}_\text{g}$ and $\text{A}_\text{u}$ representations, respectively, while the states $^3P^\text{o}(sp)$ and $^3P^\text{e}(p^2)$ (both triply degenerate) are the lowest-energy state of the $\text{B}_{k\text{u}}$ and $\text{B}_{k\text{g}}$ representation ($k=1,2,3$), respectively.
For example, the $d^2$ and $sf$ configurations are excited states of the $\text{A}_\text{g}$ and $\text{A}_\text{u}$ representation, and one can easily show that their non-interacting nodes are not exact.
This result is not surprising because we know that excited states must have additional nodes in order to be orthogonal to the lowest-energy state, and these additional nodes are usually not imposed by symmetry. \cite{QR12}

\subsubsection{Interdimensional degeneracies}
We would like to mention here that the singlet equivalent of the four triplet states for which we have found the exact expression of the nodal surface do exist. 
These are the $^1S^\text{e}(s^2)$,  $^1P^\text{o}(sp)$,  $^1D^\text{o}(pd)$ and  $^1F^\text{e}(pf)$ states. \cite{ExSpherium10}
These singlet states are connected to their triplet partner by exact interdimensional degeneracies. \cite{Herrick75a, Herrick75b, ExSpherium10}
\alert{Two states in different dimensions are said to be interdimensionally degenerated when their energies are the same.
Exploiting the relations between problems with a different number of spatial dimensions is a widespread and useful technique in physics and chemistry (see for example Ref.~\onlinecite{Doren86}).}

These types of interdimensional degeneracies also explain why the exact nodes of the $^3P^\text{e}(2p^2)$ and $^1P^\text{e}(2p^2)$ states of the helium atom are known.
Indeed, these 3D helium states are degenerate with the $^1S^\text{e}(2s^2)$ and $^3S^\text{e}(1s2s)$ states in 5D, and the exact nodes of these states are known. \cite{Bressanini05a, Bajdich05}

To illustrate this, let us take a concrete example.
In $D$ dimensions, the spatial part of the exact wave function for the $^1S^\text{e}(1s^2)$ ground state of the helium atom satisfies the following equation
\begin{equation}
	-\frac{1}{2} \Delta^{(D)} \Lambda + \left(-\frac{2}{r_1} - \frac{2}{r_2} + \frac{1}{r_{12}}\right) \Lambda= E\,\Lambda,
\end{equation}
where $\Delta^{(D)}$ is the Laplace operator in $D$ dimensions which, in terms of $r_1$, $r_2$ and $r_{12}$, reads 
\begin{equation}
\label{He-eq}
\begin{split}
	\Delta^{(D)}  
	& = \frac{\partial^2}{\partial r_1^2} + \frac{\partial^2}{\partial r_2^2}  + 2 \frac{\partial^2}{\partial r_{12}^2} 
	\\
	& + \frac{r_1^2 + r_{12}^2 - r_2^2}{r_1 r_{12}} \frac{\partial^2}{\partial r_1 \partial r_{12}}
	+ \frac{r_2^2 + r_{12}^2 - r_1^2}{r_2 r_{12}} \frac{\partial^2}{\partial r_2 \partial r_{12}} 
	\\
	& + (D-1) \left( \frac{1}{r_1} \frac{\partial}{\partial r_1} +  \frac{1}{r_2} \frac{\partial}{\partial r_2} + \frac{2}{r_{12}} \frac{\partial}{\partial r_{12}} \right).
\end{split}
\end{equation}
$\Lambda$ is a nodeless, totally symmetric function of $r_1$, $r_2$ and $r_{12}$ for any value of $D \ge 2$.
Now, let us consider the  $^3P^\text{e}(2p^2)$ state of the helium atom in $D-2$ dimensions and let us write the spatial wave function as $\Phi =  \Psi_0\,\Lambda$ where $\Psi_0 = (x_1 y_2 - y_1 x_2)$ and $\Lambda$ is a function of $r_1$, $r_2$ and $r_{12}$.
One can easily show that 
 \begin{equation}
 \begin{split}
 	\Delta^{(D)} \Phi 
	& = \Psi_0 \left[ \Delta^{(D)} \Lambda + \left( \frac{2}{r_1} \frac{\partial  \Lambda}{\partial r_1} +  \frac{2}{r_2} \frac{\partial  \Lambda}{\partial r_2} + \frac{4}{r_{12}} \frac{\partial  \Lambda}{\partial r_{12}} \right) \right]
	\\
	& = \Psi_0 \Delta^{(D+2)} \Lambda
\end{split}
\end{equation}
Therefore, $\Lambda$ satisfies Eq.~ \eqref{He-eq} and is thus a nodeless, totally symmetric function of $r_1$, $r_2$ and $r_{12}$, and the nodes are given entirely by the function  $\Psi_0 = (x_1 y_2 - y_1 x_2)$.
A similar relationship can be obtained between the  $^3S^\text{e}(1s2s)$ in 5D and the $^1P^\text{e}(2p^2)$ state in 3D.

Interdimensional degeneracies also explain why the nodes of the $^3\Sigma_\text{g}^-$ state of the H$_2$ molecule (which is degenerate with the $^1\Sigma_\text{g}^+$ state in 5D) are also known. \cite{Herrick75b, Bajdich05}
Interdimensional degeneracies could potentially be used to discover exact nodes of new atomic and molecular systems. They have been exploited very successfully in van der Waals clusters. \cite{Nightingale05}

\subsection{Three-electron systems}

\subsubsection{$^4S^{\rm o}(p^3)$ state}
The first three-electron system we wish to consider here is the $p^3$ configuration. 
It corresponds to the state where three spin-up electrons occupy the $p$ orbitals and the lowest $s$ orbital is vacant.
This state has an uniform electronic density and its non-interacting wave function is given by
\begin{equation}
	\label{p3-HF}
	\Psi_0(p^3) = 
		\begin{vmatrix}
			x_1	&	y_1	&	z_1	\\
			x_2	&	y_2	&	z_2	\\
			x_3	&	y_3	&	z_3	\\
		\end{vmatrix}
	= \bm{r}_1 \cdot (\bm{r}_{2} \times \bm{r}_{3}).
\end{equation}
In Eq.~\eqref{p3-HF}, the scalar triplet product can be interpreted as the signed volume of the parallelepiped formed by the three radius vectors $\bm{r}_1$, $\bm{r}_2$ and $\bm{r}_3$.
Thus, it is easy to understand that the non-interacting nodes of the $p^3$ configuration are encountered when the three electrons are located on a great circle, hence minimizing the volume of the parallelepiped.

For this state, we can show that the non-interacting and exact nodes are identical by using symmetry operations, as sketched in Fig.~\ref{fig:proof}.
First, we place the three electrons on a great circle which can be taken as the equator (i.e. $\theta_1 = \theta_2 = \theta_3 = \pi/2$) with no loss of generality. 
We assume that, for this configuration, the exact wave function has a value $\Psi \equiv \Psi(\{(\pi/2,\phi_1),(\pi/2,\phi_2)\},(\pi/2,\phi_3)\}) = K$.
Because this state has odd parity, inversion must change the sign of the wave function: $\Psi^\prime \equiv \Psi(\{(\pi/2,\phi_1+\pi),(\pi/2,\phi_2+\pi)\},(\pi/2,\phi_3+\pi)\}) = -K$. By applying the $C_2(z)$ rotation around the $z$ axis (which consists of adding $\pi$ to the azimuthal angle of each electron), one can bring back the electrons to their original positions. 
Due to the $S$ nature of the state, a rotation does not affect the wave function, and we have $\Psi^{\prime \prime} \equiv \Psi(\{(\pi/2,\phi_1),(\pi/2,\phi_2)\},(\pi/2,\phi_3)\}) = -K$.
Because $\Psi = \Psi^{\prime\prime}$, this means that $K = 0$.
Once again, using simple symmetry operations, we have shown that the non-interacting and exact nodes are the same.
We have confirmed this proof by performing FCI and near-exact Hylleraas calculations, and showed that the non-interacting nodes never move.
The exactness of the non-interacting nodes for this state is probably due to its high symmetry.
Moreover, the $p^3$ configuration is the lowest-energy state of $\text{A}_\text{u}$ symmetry in the $\text{D}_\text{2h}$ point group.

Let us give an alternative proof of the exactness of the non-interacting nodes for the $p^3$ configuration.
Here we will take advantage of a particular interdimensional degeneracy between a fermionic excited state and a bosonic ground state. \cite{Herrick75b}
It can be easily shown that (see Eqs.~\eqref{Phi} and \eqref{p3-HF})
\begin{equation}
	\Delta^{(3)} \Phi
	= \Delta^{(3)} \Psi_0\,\Lambda
	=  \Psi_0 \Delta^{(5)} \Lambda - \frac{6}{R^2}.
\end{equation}
Because $\Psi_0$ is antisymmetric, the condition of antisymmetry of the total wave function $\Phi$ implies that $\Lambda$ is a totally symmetric function. 
This means that $\Lambda$ is the ground-state wave function of the spinless bosonic $s^3$ state at $D = 5$.
Consequently, $\Lambda$ is nodeless and the nodes are given by the zeros of $\Psi_0$.

In the case of atomic systems, Bajdich \textit{et al.} have demonstrated that the non-interacting wave function of the $^4S(2p^3)$ state has also the same nodes as the exact wave function, \cite{Bajdich05} and this can also be attributed to a well-known interdimensional degeneracy. 
Indeed, Herrick has shown that the exact $^4S(2p^3)$ fermionic state at $D=3$ is degenerate of the spinless bosonic $1s^3$ ground state at $D = 5$. \cite{Herrick75b}

\subsubsection{$^4D^{\rm e}(sp^2)$ state}
\begin{figure}
	\includegraphics[width=0.45\textwidth]{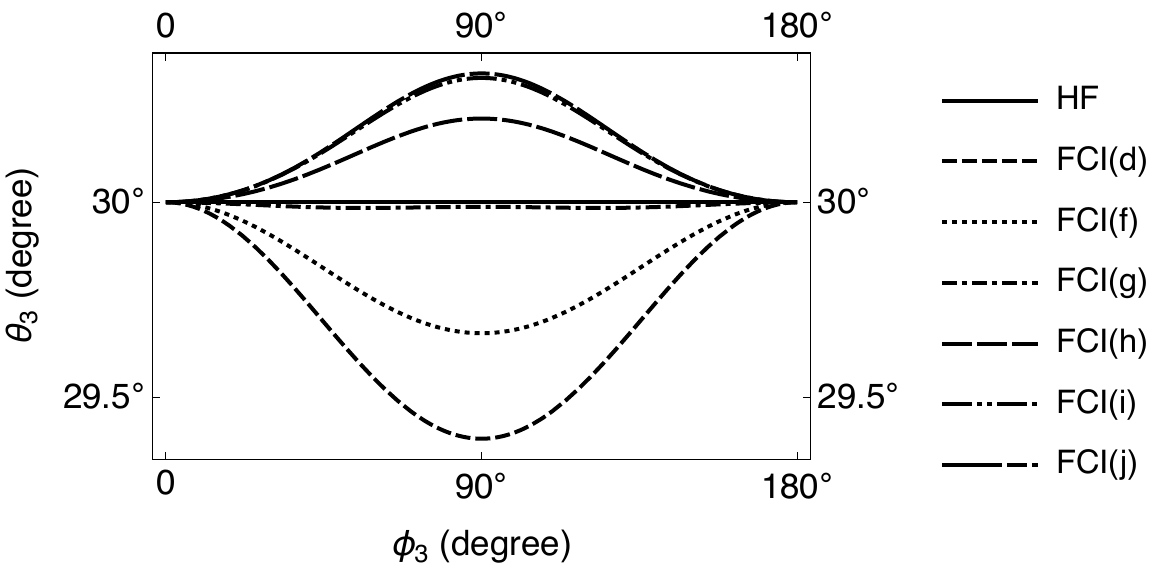}
	\caption{
	Non-interacting and FCI nodes of the $^4D^\text{e}(sp^2)$ state at $r_s = 1$. 
	The first and second electrons are at $\bm{\Omega}_1=(\pi/6,0)$ and $\bm{\Omega}_2=(\pi/6,\pi)$.
	\label{fig:sp2-nodes}
	}
\end{figure}

We now consider the quartet $D$ state created by placing one electron in the lowest $s$ orbital and two electrons in the $p$ orbitals.
This state is the ground state for three spin-up electrons on a sphere.
Unlike the $p^3$ configuration considered above, this state has a non-uniform density and its non-interacting wave function is 
\begin{equation}
	\Psi_0(sp^2) = 
		\begin{vmatrix}
			1	&	x_1	&	y_1	\\
			1	&	x_2	&	y_2	\\
			1	&	x_3	&	y_3	\\
		\end{vmatrix}
		= \bm{z} \cdot (\bm{r}_{12} \times \bm{r}_{13}).
\end{equation}
The non-interacting nodes of the $sp^2$ nodes are encountered when the three electrons are on a small circle perpendicular to the $z$ axis (see Table \ref{tab:HFwf}).
For a particular positions of the first two electrons, we have computed the FCI nodes for the $sp^2$ configuration for increasing basis set using up to $d$, $f$, $g$, $h$, $i$ and $j$ functions.
The results are reported in Fig.~\ref{fig:sp2-nodes} where we have represented the nodal surface of the $sp^2$ configuration at various level of theory and for a particular position of two electrons $\bm{\Omega}_1 = (\pi/6,0)$ and $\bm{\Omega}_2 = (\pi/6,\pi)$ for $r_s = 1$.
Based on these results, we can consider the FCI($j$) nodes as near exact.
We observe that the difference between the non-interacting and FCI nodes is always quite small (less than a degree), and that the non-interacting nodes have the same quality of a FCI($g$) calculations.
This shows that the non-interacting nodes in the $sp^2$ configuration are not identical to the nodes of the exact wave function but are nonetheless very accurate.
This state probably lacks symmetry due to the vacant $p$ orbital, and it would be interesting to know what happen in the case of the $sp^3$ configuration.

\subsection{Four-electron systems}

\subsubsection{$^5S^{\rm o}(sp^3)$ state}
The $^5S^{\rm o}(sp^3)$ state is the ground state of four spin-up electrons on a sphere and has a uniform density.
The non-interacting wave function for the $sp^3$ configuration reads 
\begin{equation}
\label{node-sp3}
	\Psi_0(sp^3) = 
	\begin{vmatrix}
		1	&	x_1	&	y_1	&	z_1	\\
		1	&	x_2	&	y_2	&	z_2	\\
		1	&	x_3	&	y_3	&	z_3	\\
		1	&	x_4	&	y_4	&	z_4	\\
	\end{vmatrix}
	= (\bm{r}_{12} + \bm{r}_{34})(\bm{r}_{12}^\times + \bm{r}_{34}^\times),
\end{equation}
and one can show that this determinant is zero if and only if the four electrons are coplanar.
This means that the non-interacting nodes of the $sp^3$ configuration corresponds to small circles.
Bajdich {\em et al.} have studied the $sp^3$ nodes in atomic systems and they have conjectured that the non-interacting nodes are \textit{``reasonably close to the exact one although the fine details of the nodal surface are not captured perfectly.''}
To the best of our knowledge, there is no known interdimensional degeneracy involving the $^5S^{\rm o}(sp^3)$ state.

To investigate further this conjecture, we computed the FCI nodes for this state for increasing basis set using up to $f$, $g$, $h$, $i$, $j$ and $k$ functions.
Because of the slow convergence of the FCI wave function, the results were inconclusive. 
However, the FCI nodes appear to converge slowly toward the non-interacting nodes, thus suggesting that the non-interacting nodes are either exact or almost exact.

To further investigate this claim, we have performed variational Monte Carlo (VMC) calculations \cite{Umrigar99} for all the states considered in this study (see Table \ref{tab:energy}).
The trial wave function that we have used for the VMC calculations is of the form $\Phi_\text{T} = \Psi_0\,e^J$ where $\Psi_0$ is given in Table \ref{tab:HFwf} and the Jastrow factor $J$ is a symmetric function of the interelectronic distances containing two-, three- and four-body terms. \cite{Huang97}
The parameters of the Jastrow factor are optimized by energy minimization. \cite{Umrigar05, Toulouse07, Umrigar07, Toulouse08}
More details will be reported elsewhere. \cite{LoosQMC} 
These VMC results are compared with benchmark calculations. 
As shown in Table \ref{tab:energy}, for all the two-electron states as well as the $p^3$ configuration for which we use the exact nodal wave function $\Psi_0$, VMC is able to reach sub-microhartree accuracy.
The same comment can be done for the $sp^3$ configuration while, for the $sp^2$ configuration where we know that $\Psi_0$ does not give a exact picture of the nodal surface, the error is more than one order of magnitude larger than for the other systems. 
This leads us to conjecture that the $sp^3$ nodes given by Eq.~\eqref{node-sp3} are identical to the nodes of the exact (yet unknown) wave function.

\begin{table}
	\caption{
	\label{tab:energy}
	VMC and benchmark energies for various states at $r_s = 1$. The statistical errors are reported in parentheses. }
		\begin{ruledtabular}
			\begin{tabular}{cll}
				States				&	VMC				&	Benchmark					\\
				\hline
				$^3P^\text{o}(sp)$		&	1.465 189 86(4)	&	1.465 189 850\footnotemark[1]		\\
				$^3P^{\rm e}(p^2)$		&	2.556 684 32(9)	&	2.556 684 316\footnotemark[1]		\\
				$^3D^{\rm e}(sd)$		&	3.556 684 32(9)	&	3.556 684 316\footnotemark[1]		\\
				$^3D^{\rm o}(pd)$		&	4.635 924 8(2)		&	4.635 924 645\footnotemark[1]		\\
				$^4S^{\rm o}(p^3)$		&	2.239 988 8(3)		&	2.239 988 9\footnotemark[1]		\\
				$^4D^{\rm e}(sp^2)$		&	1.699 883(3)		&	1.699 872\footnotemark[2]			\\
				$^5S^{\rm o}(sp^3)$		&	1.836 555 6(6)		&	1.836 556\footnotemark[2]			\\
				\end{tabular}
		\end{ruledtabular}
		\footnotetext[1]{Hyllerras-type calculation}
		\footnotetext[2]{Extrapolated FCI calculation}
\end{table}

\section{Conclusion}
In this paper, we have studied the fermionic nodes for various electronic states of the ``electrons on a sphere'' paradigm.
We have rigorously demonstrated that, for the $sp$, $p^2$, $sd$, $pd$ and $p^3$ configurations, the non-interacting wave function has the same nodes as the exact wave function.
We have shown that this peculiar feature can be attributed to exact interdimensional degeneracies. 
Interdimensional degeneracies also explain why the exact nodes of various atomic and molecular systems are known analytically. 
Therefore, we could potentially used new interdimensional degeneracies to discover the exact nodes for new atomic and molecular systems.

Even when the non-interacting nodes are not exact, we have shown that most of the features of the exact nodal surface are captured by the non-interacting nodes. 
Thus, we expect the fixed-node error to be quite small for these systems.
This could be a new, alternative way to obtain accurate near-exact energies for finite and infinite uniform electron gases.
\alert{Indeed, as illustrated in Ref.~\onlinecite{Glomium11}, the electrons-on-a-sphere model can be used to create finite and infinite uniform electron gases and we have shown that the conventional ``jellium'' model \cite{ParrBook} (i.e. electrons in a periodic box) and the present model are equivalent in the thermodynamic limit due to the the ``short-sightedness of electronic matter''. \cite{Kohn96, Kohn05}}

Although we have not been able to prove it rigorously, we have conjectured that the nodes of the non-interacting wave function for the $sp^3$ configuration are exact.
This claim is supported by numerical evidence.

\begin{acknowledgments}
P.F.L.~and D.B.~would like to thank Mike Towler for the hospitality at the TTI institute during the ``Quantum Monte Carlo in the Apuan Alps IX'' conference where this work has begun.
P.F.L.~thanks the NCI National Facility for generous grants of supercomputer time and the Australian Research Council for a Discovery Early Career Researcher Award (Grant No.~DE130101441) and a Discovery Project grant (DP140104071).
P.F.L.~thanks Peter Gill for valuable discussions. 
\end{acknowledgments}

%

\end{document}